\pdfoutput=1
\documentclass[conference]{IEEEtran}
\IEEEoverridecommandlockouts
\usepackage{times}
\usepackage{tikz}
\usetikzlibrary{positioning}
\usepackage{pgfplotstable}
\usepgfplotslibrary{fillbetween}
\usetikzlibrary{patterns}
\usepackage{subcaption}
\usepackage{subfiles}
\usepackage{xcolor}

\usepackage{cite}
\usepackage{amsmath,amssymb,amsfonts}
\usepackage{algorithmic}
\usepackage{graphicx}
\usepackage{textcomp}
\usepackage{xcolor}
\usepackage{RobStd}
\pgfmathdeclarefunction{gauss}{2}{%
  \pgfmathparse{1/(#2*sqrt(2*pi))*exp(-((x-#1)^2)/(2*#2^2))*0.7}%
}

\def \layersep{1.5cm}

\def\BibTeX{{\rm B\kern-.05em{\sc i\kern-.025em b}\kern-.08em
    T\kern-.1667em\lower.7ex\hbox{E}\kern-.125emX}}
\pgfplotsset{compat=1.14}
\begin{document}

\title{A Device Non-Ideality Resilient Approach for Mapping Neural Networks to Crossbar Arrays}

\author{ Arman Kazemi$^{*}$, Cristobal Alessandri$^\dagger$, Alan C. Seabaugh$^\dagger$, X. Sharon Hu$^{*}$, Michael Niemier$^{*}$, Siddharth Joshi$^{*}$\\
\normalsize $^{*}$Department of Computer Science and Engineering, University of Notre Dame\\
\normalsize $^\dagger$Department of Electrical Engineering, University of Notre Dame, akazemi@nd.edu
}

\vspace{-0.1in}

\maketitle
%\IEEEpeerreviewmaketitle

\begin{abstract}

We propose a technology-independent method, referred to as adjacent connection matrix (ACM), to efficiently map signed weight matrices to non-negative crossbar arrays. When compared to same-hardware-overhead mapping methods, using ACM leads to improvements of up to 20\% in training accuracy for ResNet-20 with the CIFAR-10 dataset when training with 5-bit precision crossbar arrays or lower. When compared with strategies that use two elements to represent a weight, ACM achieves comparable training accuracies, while also offering area and read energy reductions of 2.3$\times$ and 7$\times$, respectively. ACM also has a mild regularization effect that improves inference accuracy in crossbar arrays without any retraining or costly device/variation-aware training.

\end{abstract}

\section{Introduction}
Automated analysis of vast amounts of data can potentially revolutionize governance, manufacturing, medicine, and many other fields. Over the past decade, increasingly complex deep neural network (DNN) models have been proposed as a means to perform such automated analysis. The cost to train and deploy such models has grown along with model complexity, leading to a need for hardware platforms that are energy-efficient and low-latency~\cite{han2016eie}. 

A promising avenue for hardware research that addresses these challenges is based on analog domain computation of matrix-vector multiplication (MVM), a critical kernel in the forward pass of DNN training as well as inference. One family of accelerators uses analog crossbar (XBar) arrays which could be composed of emerging devices such as resistive random access memories (RRAM)~\cite{wong2012metal}, phase change memories (PCM)~\cite{Burr2017}, and ferroelectric field-effect transistors (FeFET)~\cite{jerry2018ferroelectric}, for highly parallel MVMs. A XBar array represents the input vector as analog voltages. Applying these voltages to the rows of the XBar array (where weights are stored as conductances at row/column crosspoints) induces a current along the XBar columns. The current of each column represents the dot product of the input vector and the weight vector represented by the synapse devices on the column~\cite{Burr2017}.

\begin{figure}[tbp]
    \centering
    \begin{subfigure}{0.49\columnwidth}
        \centering
        \centerline{\includegraphics[scale=0.6]{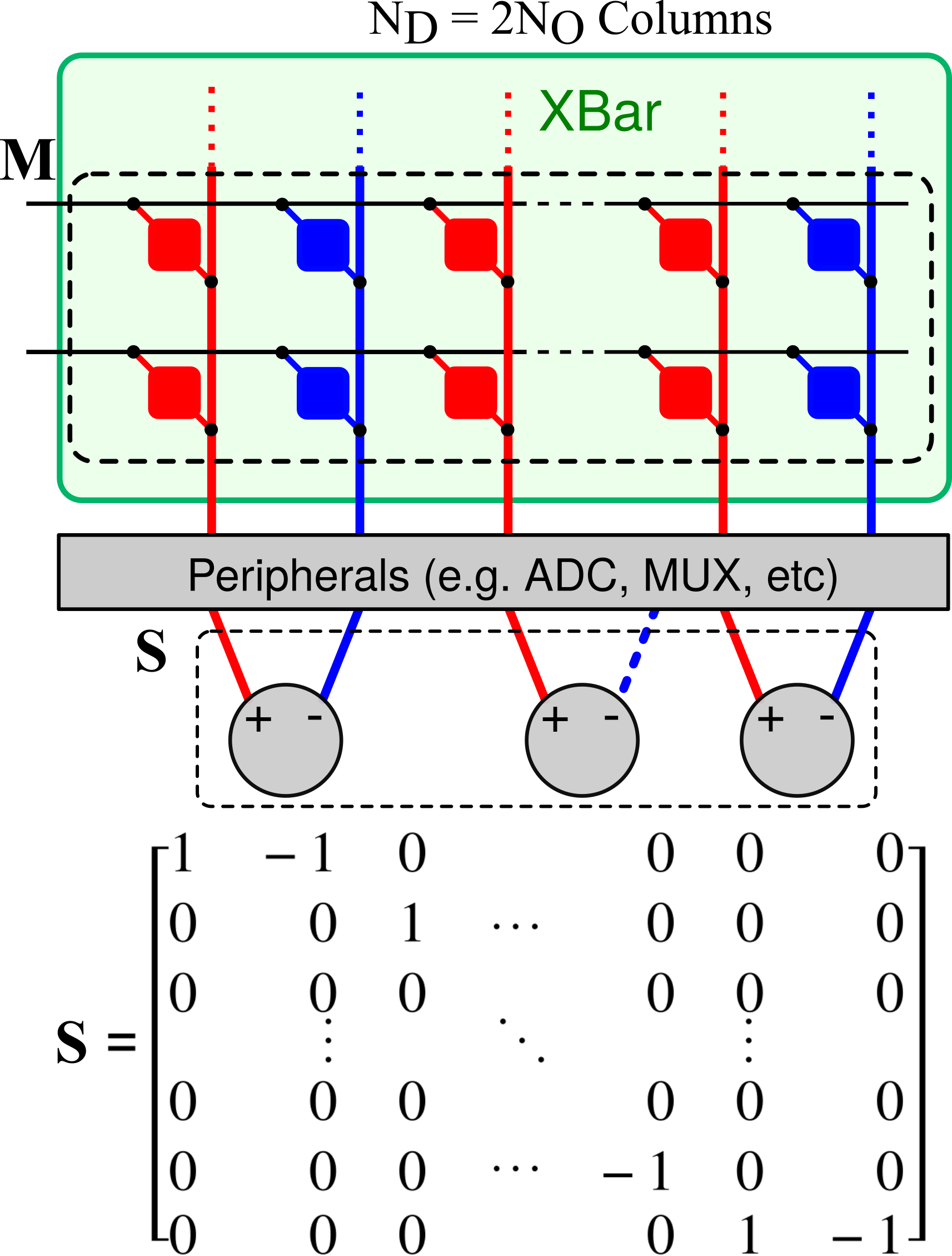}}
        \caption{Double Element (DE)}
    \end{subfigure}~
    \begin{subfigure}{0.49\columnwidth}
        \centering
        \centerline{\includegraphics[scale=0.6]{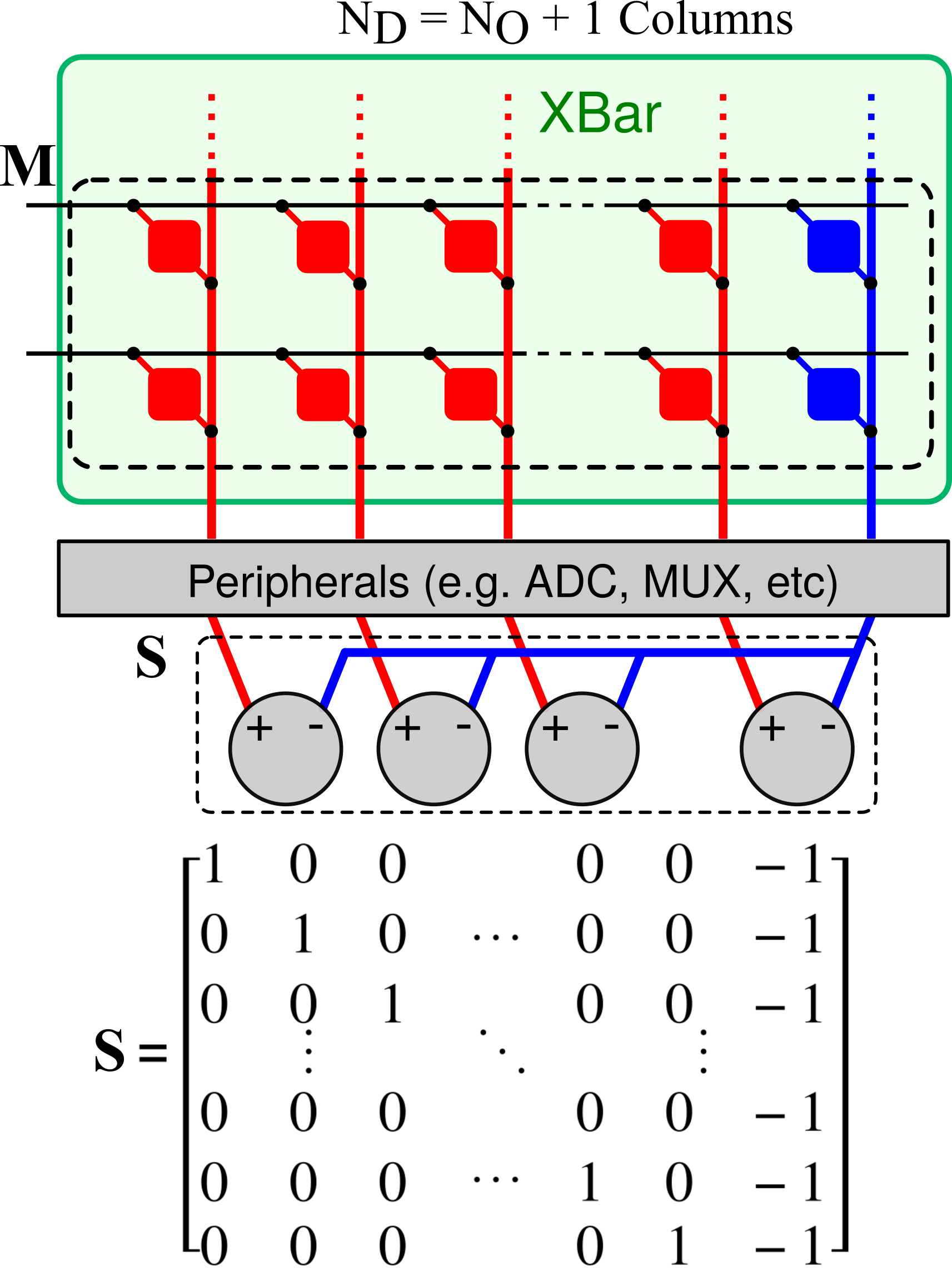}}
        \caption{Bias Column (BC)}
        % \label{fig:bc}
    \end{subfigure}
    \caption{Prior approaches to map DNN models to ReXB arrays; {\bf (a)} Using two resistive elements to represent one weight; {\bf (b)} using a single column of resistive elements as reference.}
    \label{fig:prior_mappings}
    \vspace{-0.2in}
\end{figure}

While XBar arrays efficiently implement MVMs and offer many performance advantages in energy and latency, their use poses many practical challenges. One such challenge is inherent in representing weights as conductance values. This constrains XBar arrays to use {\it non-negative} conductance values to implement arbitrary \textit{signed} MVMs. As shown in Fig.~\ref{fig:prior_mappings}, two approaches have been widely adopted: {\bf (i)} differential encoding where two elements are used to represent one weight (a {\bf double element (DE)} approach) \cite{ambrogio2018equivalent,Gokmen2016} and {\bf (ii)} a constant bias to remap values in the range $(-w,w)$ to $(0,2w)$ (a {\bf bias column (BC)} approach)~\cite{chen2017neurosim+,Chang2017}. Another challenge with performing MVMs via XBar arrays arises due to the limitations of devices employed for synapse elements, e.g., RRAMs~\cite{wong2012metal}, PCMs~\cite{Burr2017}, FeFETs~\cite{jerry2018ferroelectric}, etc, with respect to achievable weight resolution and weight update linearity, which in turn can adversely impact training accuracy. Furthermore, methods to overcome these issues, e.g., using multiple synapse elements for a single weight~\cite{chi2016prime}, further reduce the energy and area savings that might otherwise be obtained from XBar arrays. Moreover, the accuracy of models deployed on XBar arrays for inference is further degraded by device variation~\cite{chen2017neurosim+,lin2018demonstration}.

As stated in~\cite{haensch2018next}, to capitalize on the benefits offered by XBar arrays, breakthroughs in material development or architecture design are needed. Within this context, this paper presents a method, referred to as \textbf{adjacent connection matrix (ACM)}, for efficiently mapping signed MVMs to XBar arrays. By learning the most effective representation of a weight through a combination of a XBar array column and its immediate neighbor, ACM increases the effective dynamic range of weight representations. This nearest-neighbor coupling also introduces a mild regularization effect that improves resilience to device variation. ACM has been evaluated with the MNIST \cite{lecun-mnisthandwrittendigit-2010} and CIFAR-10 \cite{krizhevsky2009learning} datasets and results indicate that using ACM can lead to {\bf (i)} up to 20\% improvements in training accuracy when compared to other strategies under equivalent resource constraints (i.e., the BC approach), {\bf (ii)} comparable training accuracies coupled with area reductions of 2.3$\times$ and read energy reductions of 7$\times$ when compared to the DE approach, and {\bf (iii)} a 10\% average improvement compared to both DE and BC in the inference accuracy of a VGG network trained on the CIFAR-10 dataset with 3-bit precision XBar arrays, assuming a 15\% device variation.

The remainder of the paper is organized as follows: Section \ref{sec:bg} reviews strategies for mapping signed MVMs to XBar arrays. Section \ref{sec:acm} presents ACM and a method to evaluate and compare it with other mappings. Section \ref{sec:results} discusses the results of our simulations and quantifies DNN accuracy improvements under resource constraints and device variation; system-level evaluations of energy, area, and delay are also presented. Section \ref{sec:conclusion} concludes by summarizing our findings.

\section{Background}
\label{sec:bg}

There are two approaches typically used to map a MVM to a XBar array; these can be implemented in both the analog and the digital domain. The first approach, i.e., DE shown in Fig.~\ref{fig:prior_mappings}a, uses two XBar array columns to represent one weight column \cite{Burr2015a,Narayanan2017}. With this approach, the difference between column-pairs in the XBar array represents the equivalent signed weighted sum. The second case is an input dependent bias approach, i.e., BC \cite{chen2017neurosim+,Chang2017}. As illustrated in Fig.~\ref{fig:prior_mappings}b, a single XBar array column is used as a reference to implement the bias. The conductance of each element in this column is fixed to the middle of the conductance range. The output from this column is subtracted from the output of all other columns to compute the signed weighted sum. 

In order to perform a MVM with XBar arrays with these mapping methods, the outputs of the MVM are digitized in the periphery of the XBar array~\cite{Burr2017}. The operational overhead of the mappings are the additions and subtractions performed after digitization. Since both mappings require a single subtraction per weight, this overhead is the same for both approaches. The hardware overhead due to the number of elements used for each mapping is evaluated in Section~\ref{sec:results}.

Note that, if the conductance values of the XBar array elements are limited to the range $[G_{min},~G_{max}]$ (for simplicity, we assume $G_{min} = 0$), the weights of BC will be in the range $[-G_{max}/2,~G_{max}/2]$, with the conductance values of the bias column elements fixed to $G_{max}/2$. For DE, the range of weights will be $[-G_{max},~G_{max}]$, at the expense of using twice as many weight elements, while representing twice as many weight values as that of the BC. In short, DE utilizes 2$\times$ synapse elements and gains a 2$\times$ wider dynamic range of weight representation compared to the BC approach. However, in both the DE and BC approaches, a MVM with non-negative weights is performed in the XBar array, followed by a simple (linear) combination of the outputs from its columns to obtain an equivalent MVM with signed weights. A critical aspect of the effectiveness of these mapping solutions is the simplicity of the linear transforms implemented. These transforms consist of the addition and subtraction of values at the XBar periphery.

\section{Adjacent Connection Matrix}
\label{sec:acm}

We extend the idea of using simple linear transforms in the periphery of the XBar array circuit to map a MVM onto XBar arrays and present a new method called the adjacent connection matrix (ACM). After a brief qualitative introduction of the ACM and its benefits over other mapping methods, i.e., DE and BC, we provide a formal definition of the different mapping techniques (DE, BC, and ACM) and show how a signed MVM can be realized with a non-negative matrix representing the XBar array, and a constant signed matrix representing each mapping. Finally, we show the regularization effect of ACM using its formal definition.

\begin{figure}[tbp]
    \centering
    \centerline{\includegraphics[scale=0.6]{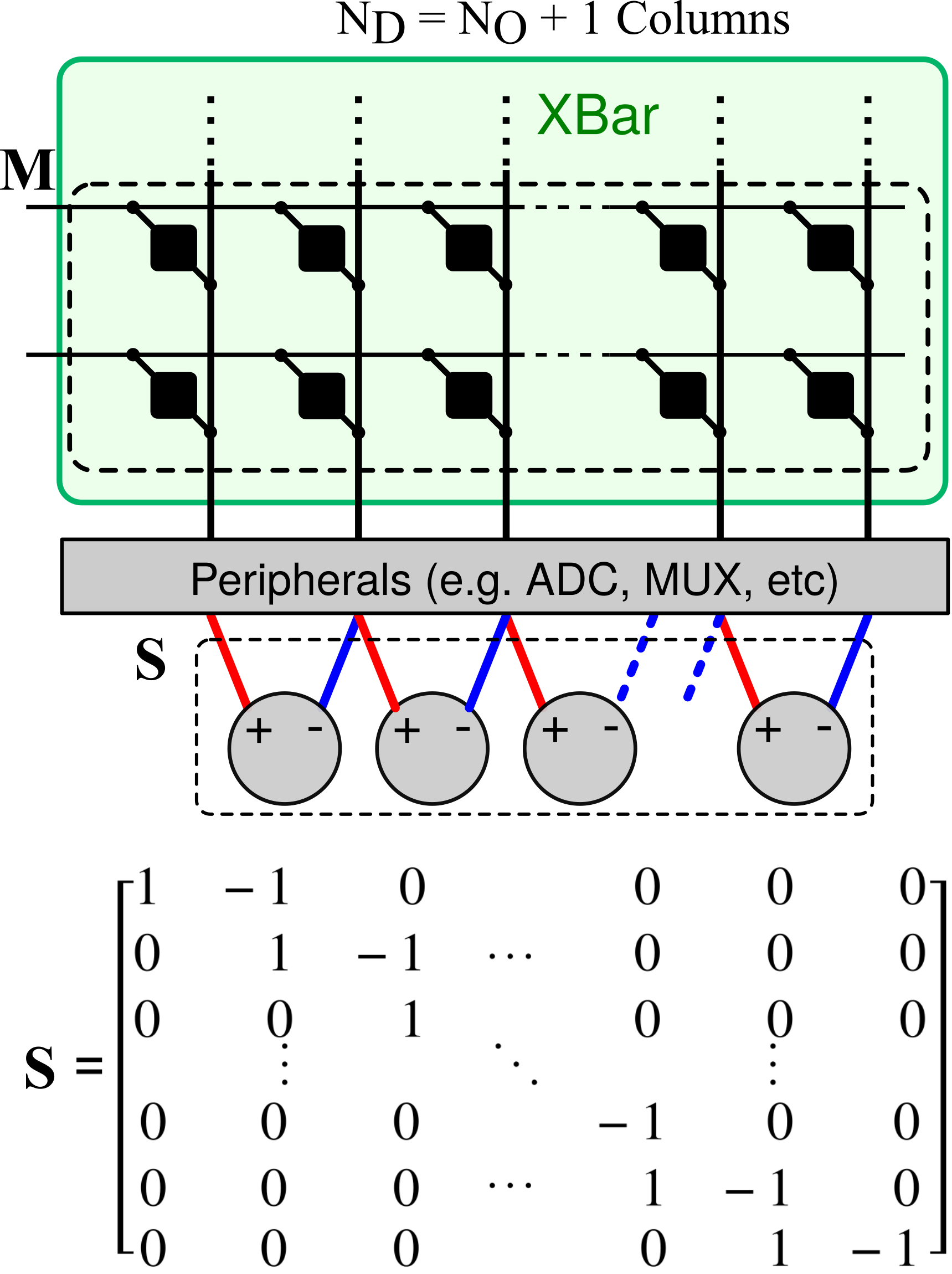}}
    \caption{ACM computes the outputs of a signed MVM as a combination of the outputs of adjacent columns with alternating signs.}
    \label{fig:acm}
    \vspace{-0.2in}
\end{figure}

\subsection{Adjacent Connection Matrix Concept}

In contrast with {\bf (i)} the DE approach that uses two elements per weight (each a reference for the other) and {\bf (ii)} the BC approach that uses a fixed reference column per MVM, the ACM approach uses each column as a reference for its immediate neighbor. ACM computes the outputs of a signed MVM as a combination of the outputs of neighboring columns of the XBar array with alternating signs as shown in Fig.~\ref{fig:acm}. Like BC, ACM also requires one additional column. This compares favorably against DE which requires two elements per weight, and therefore requires almost double the number of synapse elements in large XBar arrays. It is noteworthy that the operational overhead of ACM is the same as BC and DE, since it requires a single subtraction for each weight. Furthermore, ACM provides a mild regularization effect which increases resilience against device variation.

\begin{figure}[t]
    \centering
    
    \begin{subfigure}{.36\columnwidth}
    \vspace{0.25cm}
        \centering
        \begin{tikzpicture}[shorten >=1pt,->,draw=green, node distance=\layersep]
            \tikzstyle{every pin edge}=[<-,shorten <=1pt]
            \tikzstyle{neuron}=[circle,fill=black!25,minimum size=13pt,inner sep=0pt]
            \tikzstyle{input neuron}=[neuron, fill=gray];
            \tikzstyle{output neuron}=[neuron, fill=gray];
            \tikzstyle{hidden neuron}=[neuron, fill=white!50, draw=black];
            \tikzstyle{annot} = [align=center, text width=4em, text centered]
        
            % Draw the input layer nodes
            \foreach \name / \y in {1,...,3}
            % This is the same as writing \foreach \name / \y in {1/1,2/2,3/3,4/4}
                \path[yshift=0.0cm]
                    node[input neuron] (I-\name) at (0,-\y/1.25) {};
        
            % Draw the output layer node    
            \foreach \name / \y in {1,...,2}
                \path[yshift=-0.4cm]
                    node[output neuron] (O-\name) at (\layersep,-\y/1.25) {};    
        
            % Connect every node in the hidden layer with the output layer
            \foreach \source in {1,...,3}
                \foreach \dest in {1,...,2}
                    \path (I-\source) edge[draw=black, solid] (O-\dest);
        
            % Annotate the layers
            %\node[annot,above of=I-1, node distance=0.5cm] (hl) {X};
            % \node[annot,right of=hl] {Z};
            \node[above right=-0.15cm and 0.35cm of I-1] {\small \textbf{W}};
            \node[annot,below of=I-2] (in) {$N_I$\\nodes};
            \node[annot,right of=in] {$N_O$\\nodes};
            % \node[align=center,below=of I-1]{$N_I$\\nodes};
            % \node[align=center,below=of O-1]{$N_O$\\nodes};
        \end{tikzpicture}
        \caption{Signed Matrix \textbf{W} of a Fully-Connected Layer}
    \end{subfigure}~
    \begin{subfigure}{.6\columnwidth}
        \centering
        \begin{tikzpicture}[shorten >=1pt,->,draw=green, node distance=\layersep]
            \tikzstyle{every pin edge}=[<-,shorten <=1pt]
            \tikzstyle{neuron}=[circle,fill=black!25,minimum size=13pt,inner sep=0pt]
            \tikzstyle{input neuron}=[neuron, fill=gray];
            \tikzstyle{output neuron}=[neuron, fill=gray];
            \tikzstyle{hidden neuron}=[neuron, fill=white!50, draw=black];
            \tikzstyle{annot} = [align=center,text width=4em, text centered]
        
            % Draw the input layer nodes
            \foreach \name / \y in {1,...,3}
            % This is the same as writing \foreach \name / \y in {1/1,2/2,3/3,4/4}
                \path[yshift=0.0cm]
                    node[input neuron] (I-\name) at (0,-\y/1.25) {};
        
            % Draw the hidden layer nodes
            \foreach \name / \y in {1,...,3}
                \path[yshift=0.0cm]
                    node[hidden neuron] (H-\name) at (\layersep,-\y/1.25) {};

            \foreach \name / \y in {1,...,2}
                \path[yshift=-0.4cm]
                    node[output neuron] (O-\name) at (\layersep+\layersep,-\y/1.25) {};    
        
            % Connect every node in the input layer with every node in the
            % hidden layer.
            \foreach \source in {1,...,3}
                \foreach \dest in {1,...,3}
                    \path (I-\source) edge (H-\dest);
        
            % Connect every node in the hidden layer with the output layer
            \foreach \source in {1,...,3}
                \foreach \dest in {1,...,2}
                    \path (H-\source) edge[draw=black, dashed] (O-\dest);
        
            % Annotate the layers
            \node[annot,above of=H-1, node distance=0.5cm] (hl) {\small dummy};
            % \node[annot,left of=hl] {\small X};
            % \node[annot,right of=hl] {\small Z};
            \node[above left=-0.1cm and 0.35cm of H-1] {\small \textbf{M}};
            \node[above right=-0.1cm and 0.35cm of H-1] {\small \textbf{S}};
            \node[annot,below of=I-2] (in) {\small $N_I$\\nodes};
            \node[annot,right of=in] (dum) {\small $N_D$\\nodes};
            \node[annot,right of=dum] {\small $N_O$\\nodes};
        \end{tikzpicture}
        \caption{Equivalent Non-negative Matrix \textbf{M} and the Periphery Matrix \textbf{S}}
    \end{subfigure}
    \caption{An example in the context of fully-connected layers, where the original matrix is decomposed as a sequence of a non-negative matrix \textbf{M}, followed by a periphery matrix \textbf{S}.}
    \label{fig:connection_matrix_diagram}
    \vspace{-0.2in}
\end{figure}
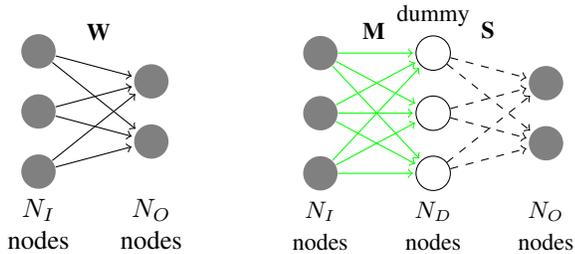

\subsection{Formal Definition of Different Mappings}\label{sec:mappings}

ACM, DE, and BC, all combine the outputs from columns of the XBar array in a fixed and predefined way. This combination of columns is comprised of additions and subtractions only and can be thought of as a matrix with its non-zero entries limited to $\pm1$. We refer to this matrix as the periphery matrix. The three different mappings and their corresponding unique periphery matrices are presented in Figs.~\ref{fig:prior_mappings} and \ref{fig:acm}. In order to verify that the periphery matrix holds as a general technique to map MVMs onto XBar arrays, we first demonstrate the decomposition of a signed MVM into {\bf (i)} a non-negative matrix that is stored on the XBar array and {\bf (ii)} a fixed, signed matrix. We then characterize the requirements of this matrix and demonstrate that it can be implemented through addition and subtraction operations at the periphery of the XBar array.

Consider an arbitrary signed matrix \textbf{W} with dimensions $N_{O}\times N_{I}$\footnote{In this section we examine the transpose of the matrices to simplify the solution of equations and the explanations.}. To constrain the multiplication to non-negative weights only, matrix \textbf{W} is factored into a matrix with non-negative elements \textbf{M}, followed by a periphery matrix \textbf{S}, i.e.,
\begin{equation}\label{eq:factor}
    \boldsymbol{S\thinspace M} = \boldsymbol{W},\;\;\boldsymbol{M}\geq0,
\end{equation}
where \textbf{M}$\geq0$ means that all elements of \textbf{M} are non-negative. Within the context of DNN layers, \textbf{W} is the weight matrix of a fully-connected layer with $N_{I}$ inputs and $N_{O}$ outputs as shown in Fig.~\ref{fig:connection_matrix_diagram}a. For ease of visualization, we define a \textit{dummy layer} $Y$ with $N_D$ neurons. Fig.~\ref{fig:connection_matrix_diagram}b shows the layer mapped to a XBar array where \textbf{M} is the non-negative matrix with dimensions $N_D\times N_I$, and \textbf{S} is a periphery matrix of dimensions $N_O \times N_D$. While we have used a fully-connected layer from a DNN in our example, all linear transforms, including convolutions, are possible through ACM.

There are two properties we desire from \textbf{S}: {\bf (i)} a fixed \textbf{S} must guarantee that any multiplication using a signed matrix \textbf{W} can be realized using a non-negative matrix \textbf{M} and a fixed signed matrix \textbf{S} and {\bf (ii)} \textbf{S} must be of a form that does not impose large hardware implementation costs.

\subsection{Sufficient Conditions of a Periphery Matrix}

As before, given a \textbf{W} with the dimensions $N_O \times N_I$, \textbf{S} can be assigned dimensions $N_O \times N_D$, and \textbf{M} the dimensions $ N_D \times N_I $. Formulated independently for each column of \textbf{W} and \textbf{M}, this can be expressed as:
\begin{equation}
    \boldsymbol{S\thinspace m_k} = \boldsymbol{w_k},\;\;\boldsymbol{m_k}\geq0,
    \label{eq:problem_def}
\end{equation}
where $\boldsymbol{m_k}$ and $\boldsymbol{w_k}$ are the $k$-th columns of \textbf{M} and \textbf{W}, respectively, and $k \in \{1, 2,\ldots ,N_{I}\}$.

A necessary condition for the existence of a solution to Eq.~\eqref{eq:problem_def} is that $\boldsymbol{w_k}$ is in the column space of \textbf{S}. This condition will be satisfied for any arbitrary $\boldsymbol{w_k}$ if and only if $\mathrm{rank}(\boldsymbol{S}) = N_{O}$. The sufficient condition for the existence of a non-negative solution is the existence of a vector $x_h$ in the null space of \textbf{S} with strictly positive elements. This guarantees that any particular solution $\boldsymbol{x_p}$ to the system $\boldsymbol{S\thinspace m_k} = \boldsymbol{w_k}$ can be shifted as $\boldsymbol{x_p'}=\boldsymbol{x_p}+\alpha \boldsymbol{x_h}$ to be non-negative. The sufficient conditions are summarized as:
\begin{eqnarray}
    1. &\mathrm{rank}(\boldsymbol{S})=N_{O} \nonumber \\
    2. &\exists \;\boldsymbol{x_h} > 0, \boldsymbol{x_h} \in \mathrm{R}^{N_{D}},\; \mathrm{s.t.}\; \boldsymbol{S\thinspace x_h} = 0. 
    \label{eq:Sconditions}
\end{eqnarray}

If these conditions are met, the signed matrix \textbf{W} can be decomposed to a non-negative matrix \textbf{M} and periphery matrix \textbf{S}, such that, $\boldsymbol{W} = \boldsymbol{S}\thinspace \boldsymbol{M}$. Equations $N_D = \mathrm{rank}(\boldsymbol{S})+\mathrm{nullity}(\boldsymbol{S})$ and $\mathrm{nullity}(\boldsymbol{S}) \geq 1$ hold, if there is at least one element ($\boldsymbol{x_h}$) in the null space of \textbf{S}. Therefore, \textbf{M} with $N_D$ columns has at least one more column than \textbf{W} with $N_O$ columns. A particular case that satisfies the second condition of Eq.~\eqref{eq:Sconditions} is $x_h=\mathbf{1}$, which implies that the elements of the rows of \textbf{S} add up to 0. Thus, an \textbf{S} such that neighboring columns are subtracted from each other, introduced earlier as the ACM, satisfies both conditions, i.e., one extra column in \textbf{M} and the sum of rows of $\boldsymbol{S} = 0$. Note that \textbf{S} also has all the properties listed as desirable per Section~\ref{sec:mappings}.

\subsection{Analysis of Different Mappings}

Using the periphery matrix decomposition discussed above, we can derive not only the ACM mapping but also the DE and BC mappings. We can observe in Figs.~\ref{fig:prior_mappings} and \ref{fig:acm} that all three approaches satisfy the conditions stated in Eq.~\eqref{eq:Sconditions}. In all cases, each row in the periphery matrix has two nonzero elements (1 and $-1$), hence the sum of the elements of the rows add up to 0. Furthermore, $N_D \geq 1$ is true for all three cases. However, DE has $N_D = 2N_O$ columns, whereas BC and ACM have the minimum number of columns, $N_D = N_O+1$. Therefore, BC and ACM require minimal additional hardware resources. Furthermore, we assume that the elements of \textbf{M} have a conductance range of $[G_{min},\, G_{max}]$ (again, for simplicity, we assume $G_{min} = 0$). Thus, by using the ACM approach, addition and subtraction of neighboring elements can result in the representation of weights over the range $[-G_{max},\, G_{max}]$ while using the same hardware resources as BC. That said, while DE can always demonstrate the full range in weights, ACM is limited by having to balance DNN accuracy and weight range, as neighboring columns are not guaranteed to have a large disparity in weights. Section~\ref{sec:results} quantifies the effect of these different mappings on system-level DNN training and inference accuracy in the presence of non-ideal XBar array synapse devices.

\subsection{Regularization Effect of Adjacent Connection Matrix}
\label{sec:regularization}
An observation of the nearest-neighbor coupling induced by the ACM approach naturally leads to the question, how does the ACM approach constrain the neighboring weights and what is the consequence of such constraint? In this subsection we examine the effects of these constraints through the lens of \emph{regularization}. Let us denote the sum of all the elements of a column of \textbf{M} as $\boldsymbol{M_j}$, in other words $\boldsymbol{M_j} = \sum_{i=1}^{N_I} \boldsymbol{M_{i j}}$. Inserting this expansion into Eq.~\eqref{eq:factor} and explicitly writing out the values in the periphery matrix \textbf{S} of ACM (Fig.~\ref{fig:acm}) leads to the following expression:
\vspace{-0.08in}

\begin{align}\label{eq:constraint}
    \begin{split}
    \sum_{i=1}^{N_I} \sum_{j=1}^{N_O} \boldsymbol{W_{ij}} ={}& \boldsymbol{M_1} - \boldsymbol{M_2} + \boldsymbol{M_2} - \boldsymbol{M_3} + \dots \\
    & + \boldsymbol{M_{N_{O-1}}} - \boldsymbol{M_{N_O}}  
    \end{split}\nonumber\\ 
    ={}& \boldsymbol{M_1} - \boldsymbol{M_{N_O}}.
\end{align}
Eq.~\eqref{eq:constraint} demonstrates that any non-negative weight matrix \textbf{M} trained with the ACM approach must satisfy the constraints on the first and last columns in the matrix \textbf{M}. For a quantized matrix where the elements of matrix \textbf{M} have bit precision B, each element can only be assigned one of $2^B$ values. Thus, for any index $j$, the column $\boldsymbol{M_j}$ can be assigned 1 of $2^B$ values per element, leading to $N_I\times 2^B$ different values. Consequently, for quantized matrices, Eq.~\eqref{eq:constraint} constrains $\sum_{i=1}^{N_I} \sum_{j=1}^{N_O} \boldsymbol{W_{ij}}$ to $2 \times N_I \times 2^B - 1$ values. When each element in \textbf{M} has a small set of possible values (i.e., when $2^B$ is smaller), this constraint is tighter. Thus, leading to a regularization effect when training with ACM. As the results in Section~\ref{sec:variation} indicate, the nearest-neighbor coupling of ACM increases device variation resilience in reverse relation with $B$ (the smaller $B$, the higher the resilience). However, ACM based training is not meant to replace standard regularization methods, e.g. L-2, dropout, etc, which have a much stronger regularization effect.

\section{Evaluation and Results}
\label{sec:results}

\begin{figure}[tp]
    \centering
    \begin{subfigure}[t]{.49\columnwidth}
        \begin{tikzpicture}
            \begin{axis}
            [
                height=80,width=0.8\columnwidth,scale only axis,
                %tick style={draw=none},
                grid=none,     
                xmin=-1,
                xmax=22,
                axis x line=bottom,
                xlabel={Pulse Number},
                ylabel={Conductance (S)},
                ylabel style={rotate=-90},
                ymax=1.05,
                ymin=0,
                axis y line=middle,
                ylabel near ticks,
                yticklabels={,,},
                xticklabels={,,},
                legend image post style={scale=0.5},
                legend style={at={(0.22,0.)},anchor=south west,fill=none,font=\footnotesize},
                every axis plot/.append style={ultra thick},
                ticks=none
            ]
            \node (b) at (2,.9) {(a)};
                \addplot%
                [
                    blue,%
                    draw=none,
                    mark=*,
                    samples=25,
                    domain=0.3:10.3,
                ]
                (x,{1/(1+exp(-x+5.3))*0.95 + 0.05});
                \addplot%
                [
                    red,%
                    draw=none,
                    mark=*,
                    samples=25,
                    domain=10.3:20.3,
                ]
                (-x+30.6,{1/(1+exp(-x+15.6))*0.95 + 0.05});
                \legend{Potentiation,Depression};
                
            \end{axis}
        \end{tikzpicture}
        %\caption*{a) Up/down symmetric non-linearity observed in \cite{jerry2018ferroelectric} and assumed in training with non-linear weight update}
    \end{subfigure}~
    \begin{subfigure}[t]{.49\columnwidth}
        \begin{tikzpicture}
            \begin{axis}[
              no markers,
              height=80,width=0.8\columnwidth,scale only axis,
            %  domain=-6:6,
              samples=100,
              axis x line=bottom,
              axis y line=middle,
              ymax=2,
              ylabel={PDF},
              xlabel={Conductance (S)},
              ylabel near ticks,
              yticklabels={,,},
              xticklabels={,,},
              enlargelimits=false, clip=false, axis on top,
              every axis plot/.append style={ultra thick},
              ticks=none,
              legend style={at={(0.85,.95)},fill=none,font=\footnotesize},
            ]
            \node (b) at (0.7,1.7) {(b)}; % label
            \node (x) at (1.3,0.5) {};
            \node (y) at (1.8,0.5) {};
            \draw[<->] (x) -- node[below] {$\sigma$} (y);
            \addplot [very thick,blue!,domain=0.4:2.4] {gauss(1.4,.25)};
            \addlegendentry{State 1};
            %\addplot [very thick,red!,domain=1:3] {gauss(2,.2)};
            \node (x1) at (2.7,0.5) {};
            \node (y1) at (2.2,.5) {};
            \draw[<->] (x1) -- node[below] {$\sigma$} (y1);            
            \addplot [very thick,red!,domain=1.6:3.6] {gauss(2.6,.25)};
            \addlegendentry{State 2};
            \draw [dashed] (1.4,0) -- (1.4,1.1);
            \draw [dashed] (1.7,0) -- (1.7,0.5);
            \draw [dashed] (2.6,0) -- (2.6,1.1);
            \draw [dashed] (2.3,0) -- (2.3,0.5);
            %\addplot [very thick,yellow!,domain=3:5] {gauss(4,.2)};
            \end{axis}
            \end{tikzpicture}
            %\caption*{b) }
    \end{subfigure}
    \caption{{\bf (a)} Up/down symmetric non-linearity observed in \cite{jerry2018ferroelectric,woo2018resistive} and assumed in training with non-linear weight update; {\bf (b)} Gaussian distribution used to model device variation for a 1-bit (2 state) device, observed in \cite{lin2018demonstration}. }
    \label{fig:nonlinearity}
    \vspace{-0.2in}
\end{figure}
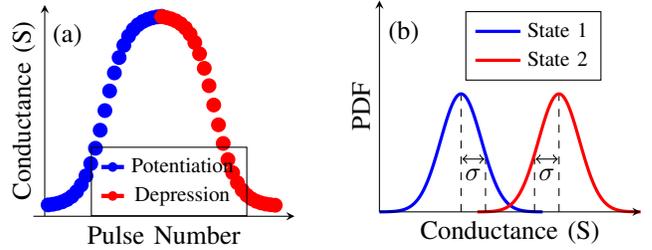

In this section, we first evaluate the training accuracy of the ACM method and compare it with alternative mappings. We follow this with an evaluation of inference accuracy on a pre-trained network with different mappings when the weights are subject to device variation. We have developed a model for neural network training using TensorFlow \cite{abadi2016tensorflow} that incorporates the non-idealities of the synapse devices. While training, matrix \textbf{M} is constrained to be non-negative and is followed by a periphery matrix that is defined as a fixed layer with values in $\{-1, +1, 0\}$ as depicted in Figs.~\ref{fig:prior_mappings} and \ref{fig:acm}. In our studies, we consider two non-ideal device characteristics that virtually exist in all physical synapse devices used for XBar arrays and impact the accuracy of DNNs trained on XBar arrays: {\bf (i)} limited weight precision (i.e., the number of representable states) and {\bf (ii)} non-linear weight update of XBar array synapse devices. For the former, we quantize the weights similar to \cite{zhou2016dorefa}, and for the latter, we present results for devices with symmetric increase/decrease steps~\cite{jerry2018ferroelectric,woo2018resistive} (Fig.~\ref{fig:nonlinearity}a) to isolate the effect of the nonlinear weight update on ACM from the effect of the nonlinearity on the learning rule. Since ACM is a linear transform, it is also compatible with learning rules tailored for devices with asymmetric weight update nonlinearity~\cite{fouda2018independent}. We present results for activations quantized to 8 bits of resolution; results on 6-bit quantized activations followed the same trend and were omitted for brevity.

To evaluate DNN training accuracy when using ACM, we train a variant of LeNet~\cite{lecun2015lenet} with the MNIST dataset; we also train a VGG-9~\cite{simonyan2014very} network (with 6 convolutional layers and 3 fully-connected layers) and a ResNet20 network~\cite{he2016deep} with the CIFAR-10 dataset using a vanilla stochastic gradient descent. We train four types of models for the above networks: {\bf (i)} a baseline model, i.e., the original network trained with signed weights, {\bf (ii)} DE, i.e., a network trained using non-negative weights and the periphery matrix in Fig.~\ref{fig:prior_mappings}a, {\bf (iii)} BC, i.e., a network trained using non-negative weights and the periphery matrix in Fig.~\ref{fig:prior_mappings}b, and {\bf (iv)} ACM, i.e., the network trained using non-negative weights and the periphery matrix in Fig.~\ref{fig:acm}. We also evaluate the impact of device variation on the inference accuracy of the VGG-9 network trained with the CIFAR-10 dataset when different mappings are used. Variation is modeled as a zero-mean, normal distribution~\cite{lin2018demonstration} as depicted in Fig.~\ref{fig:nonlinearity}b, and this is added to the desired conductance value. Finally, we investigate the system-level characteristics of a XBar-based accelerator that uses different mappings with the NeuroSim+~\cite{chen2017neurosim+} tool in terms of area, delay, and energy.

\subsection{Neural Network Training Accuracy}
\label{sec:training}

\subfile{./Figs/test}

Figs.~\ref{fig:results}a and~\ref{fig:results}e show training and test accuracies for the LeNet and ResNet20 networks with single-precision floating-point (FP32) weights, respectively. The three mapping schemes achieve results equivalent to the baseline model. Furthermore, the training and test errors follow similar trajectories as a function of the number of epochs. This is consistently observed for different networks and training conditions for the two datasets. These observations validate our analysis in Section~\ref{sec:acm}, experimentally showing that the three decompositions (DE, BC, ACM) can achieve similar accuracy in DNN training when there are no constraints on weight precision during training. Note that while ACM provides identical test accuracy at FP32 weights, the training accuracy is lower than that of DE and BC in Fig.~\ref{fig:results}e. This is due to the mild regularization effect of ACM that is discussed in Section~\ref{sec:regularization}.

Figs.~\ref{fig:results}b,~\ref{fig:results}c, and~\ref{fig:results}d show the effect of limited XBar array weight precision on training DNNs on MNIST and CIFAR-10 tasks. The test errors are shown as a function of the number of bits for weight precision. Since there exists no {\em array-level} experiments demonstrating XBar array synapse devices with weight precision higher than 5 bits, we focus our study on weights from 2-6 bits. For precision lower than 6 bits, it can be observed that the error of DE is lower than that of the other mappings. As one would expect, this is because DE uses twice the number of elements as BC and ACM and consequently has twice the range in weight representation. When using ACM, some of the resolution lost through BC is recovered, placing its accuracy between BC and DE. The ACM encoding distributes the weight value over two columns to provide better tolerance to limited resolution compared to the BC approach. Thus, at {\em resource parity}, ACM provides a resolution advantage over the BC approach. At precision higher than 5 bits, for DNNs trained on the MNIST dataset, training accuracy saturates to values achieved through FP32 training. However, results on the CIFAR-10 dataset (Figs.~\ref{fig:results}c and~\ref{fig:results}d) start to diverge from the FP32 results for precision lower than 8 bits due to the higher complexity of the task.

When training with non-linear weight update (Figs.~\ref{fig:results}f,~\ref{fig:results}g, and~\ref{fig:results}h), the differences between DE and BC approaches become even more apparent. As before, the error when training with ACM is lower than the error obtained when training using BC, approaching the error of DE. However, the accuracy improvement is much more dramatic due to the disparate accuracy impact of the nonlinear weight update. These results show that ACM consistently improves upon the accuracy of BC while using the same hardware resources. The VGG-9 network is overparameterized and better offsets the non-linearity as compared to the ResNet20 network. Therefore, the accuracy decline starts at 5 bits in Fig.~\ref{fig:results}g rather than 6 bits in Fig.~\ref{fig:results}h. The largest gain is obtained for non-linear weights when training with 5-bit precision XBar arrays or lower: e.g. for ResNet20, effectively 2 bits in weight resolution are recovered, leading to a 20\% improvement in accuracy.

\subsection{Effects of Device Variation on Neural Network Inference}
\label{sec:variation}
\begin{figure}[tbp]
    \centering
    %\vspace{-0.07in}
    \begin{subfigure}{0.49\columnwidth}
        \begin{tikzpicture}%[scale=0.5]
            \begin{axis}[
                    scale=0.5,
                    title=1-bit Weights,
                    ybar=.7pt,
                    enlargelimits=0.15,
                    bar width = .10cm,
                    ylabel={Accuracy (\%)},
                    symbolic x coords={0,5,10,15,20,25},
                    xtick=data,
                    % xticklabels=\empty,
                    % xtick pos=bottom,
                    legend image post style={scale=0.5},
                    axis line style=ultra thick,
                    every tick/.style={     % make ticks thick and black
                    black,
                    thick,
                    },
                    legend style={font=\footnotesize,
                                fill=none, at={(0.621,0.51)},anchor=south west},
                    title style={yshift=-1.5ex,},
                    ]
                \addplot[fill=gold] coordinates {(0, 83.5) (5, 81.81079999999997) (10, 77.02799999999998) (15, 66.25319999999999) (20, 46.24000000000001) (25, 31.827999999999996)};
                \addplot[fill=blue] coordinates {(0, 82.72000000000001) (5, 81.98360000000001) (10, 79.3076) (15, 74.48200000000001) (20, 64.50479999999999) (25, 49.4024)};
                \addplot[fill=red] coordinates {(0, 78.75) (5, 74.19599999999998) (10, 59.218399999999995) (15, 38.1156) (20, 23.796799999999998) (25, 15.766000000000002)};
                \legend{DE,ACM,BC}
            \end{axis}
        \end{tikzpicture}
    \end{subfigure}~
    \begin{subfigure}{0.49\columnwidth}
        \begin{tikzpicture}%[scale=0.5]
            \begin{axis}[
                    scale=0.5,
                    title=3-bit Weights,
                    ybar=.7pt,
                    enlargelimits=0.15,
                    bar width = .10cm,
                    axis line style=ultra thick,
                    every tick/.style={     % make ticks thick and black
                    black,
                    thick,
                    },
                    symbolic x coords={0,5,10,15,20,25},
                    xtick=data,
                    % xticklabels=\empty,
                    % xtick pos=bottom,
                    title style={yshift=-1.5ex,},
                    ]
                \addplot[fill=gold] coordinates {(0, 90.05999999999996) (5, 89.6912) (10, 88.0996) (15, 84.72160000000002) (20, 78.66680000000001) (25, 68.5256)};
                \addplot[fill=blue] coordinates {(0, 89.47) (5, 89.11999999999998) (10, 87.8212) (15, 85.44800000000001) (20, 80.40119999999999) (25, 70.0076)};
                \addplot[fill=red] coordinates {(0, 87.64000000000004) (5, 86.0704) (10, 78.35960000000001) (15, 64.622) (20, 46.91600000000001) (25, 27.2696)};
                %\legend{ACM,DE,BC}
            \end{axis}
        \end{tikzpicture}
    \end{subfigure}
    \\
    \begin{subfigure}{0.49\columnwidth}
        \begin{tikzpicture}%[scale=0.5]
            \begin{axis}[
                    scale=0.5,
                    title=4-bit Weights,
                    ybar=.7pt,
                    enlargelimits=0.15,
                    bar width = .10cm,
                    axis line style=ultra thick,
                    every tick/.style={     % make ticks thick and black
                    black,
                    thick,
                    },
                    ylabel={Accuracy (\%)},
                    xlabel={Sigma of Variation (\%)},
                    symbolic x coords={0,5,10,15,20,25},
                    xtick=data,
                    % xtick pos=bottom,
                    title style={yshift=-1.5ex,},
                    ]
                \addplot[fill=gold] coordinates {(0, 91.11999999999995) (5, 90.94800000000001) (10, 90.27599999999998) (15, 89.07799999999999) (20, 86.9676) (25, 83.44760000000001)};
                \addplot[fill=blue] coordinates {(0, 91.03999999999998) (5, 90.7156) (10, 89.6892) (15, 87.8888) (20, 84.51760000000002) (25, 78.1968)};
                \addplot[fill=red] coordinates {(0, 90.28999999999998) (5, 89.0404) (10, 85.8832) (15, 75.6344) (20, 62.696000000000005) (25, 44.57920000000001)};
                %\legend{ACM,DE,BC}
            \end{axis}
        \end{tikzpicture}
    \end{subfigure}~
    \begin{subfigure}{0.49\columnwidth}
        \begin{tikzpicture}%[scale=0.5]
            \begin{axis}[
                    scale=0.5,
                    title=6-bit Weights,
                    ybar=.7pt,
                    enlargelimits=0.15,
                    axis line style=ultra thick,
                    every tick/.style={     % make ticks thick and black
                    black,
                    thick,
                    },
                    bar width = .10cm,
                    legend style={at={(0.5,-0.15)},
                        anchor=north,legend columns=-1},
                    xlabel={Sigma of Variation (\%)},
                    symbolic x coords={0,5,10,15,20,25},
                    xtick=data,
                    % xtick pos=bottom,
                    title style={yshift=-1.5ex,},
                    ]
                \addplot[fill=gold] coordinates {(0, 92.35999999999996) (5, 92.192) (10, 91.85) (15, 91.30039999999997) (20, 90.4524) (25, 89.15960000000001)};
                \addplot[fill=blue] coordinates {(0, 92.09999999999997) (5, 91.79679999999999) (10, 91.11280000000001) (15, 88.95920000000002) (20, 86.40639999999998) (25, 80.458)};
                \addplot[fill=red] coordinates {(0, 91.99) (5, 91.1684) (10, 89.16279999999999) (15, 86.34360000000002) (20, 79.7832) (25, 70.60439999999998)};
                % \legend{ACM,DE,BC}
            \end{axis}
        \end{tikzpicture}
    \end{subfigure}
    \caption{Effects of device variation on the inference accuracy of a VGG-like network trained with different mapping approaches and bit precision on the CIFAR-10 dataset}
    \label{fig:variation}
    \vspace{-0.2in}
\end{figure}
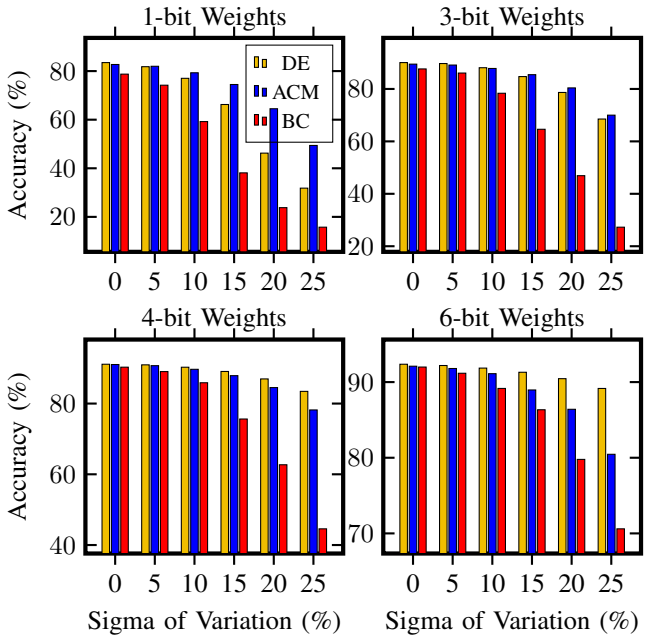

We evaluate the inference accuracy of a VGG-9 network trained on CIFAR-10 under the conditions of device variation when operating with different weight precision. After training, variation is added to the trained model weights and the inference accuracy is evaluated without any fine-tuning. Fig.~\ref{fig:variation} shows the results averaged over 25 samples per data point for 4 different precision values. There is a disparity in DNN inference accuracy even when not subjected to any device variation (e.g., see Fig.~\ref{fig:results}e for the difference in quantized DNNs trained on CIFAR-10 with DE, BC, and ACM). Results on inference with added variation show that this initial disparity is dramatically increased and BC consistently performs worse than the other two mappings regardless of the bit precision. Due to limited space, we only show results for 1-bit, 3-bit, 4-bit, and-6 bit weights; the 2-bit and 5-bit weights follow the expected trends. ACM performs better than DE and BC for 1, 2, and 3-bit weights and DE outperforms the other mapping methods for the 4, 5, and 6-bit weights. The improvement in inference accuracy when using ACM at lower bit precision may appear counterintuitive. However, this can be explained by considering the regularization effect of ACM discussed in Section~\ref{sec:regularization}. At lower bit precision, the constraint is tighter, while it is relaxed at higher bit precision. This constraint strengthens the network against device variation.

\subsection{System-Level Evaluation}

Table~\ref{tab:sys} shows system-level results for the three mapping approaches generated using the NeuroSim+~\cite{chen2017neurosim+} tool with the default parameters in the 14nm tech node. The assumed peripherals include MUX, ADC, word-line decoder, bit-line and select-line switch matrices, adders, and shift registers. The read energy and latency values are for one epoch of training a two-layered multi-layer perceptron (MLP) network with a XBar-based hardware accelerator. Read energy, area, and read delay values for BC and ACM approaches are exactly the same, as there is practically no difference in their hardware resource utilization. The read energy of DE is 7$\times$ more than that of the ACM due to the longer wires for rows of the XBar array. DE uses 2.3$\times$ XBar area compared to the ACM, as it requires twice as many elements. The peripherals are also larger and require more area. Furthermore, DE has a 1.33$\times$ higher read delay due to the additional columns that need to be multiplexed for the associated peripherals.

\section{Conclusion}
\label{sec:conclusion}

We introduced the ACM mapping method to mitigate the effects of limited weight resolution and non-linearity on neural network training while incurring minimal hardware overhead. We demonstrated, both mathematically and by simulation, that ACM is a general approach and represents the same MVMs as previous mapping approaches. Neural network training evaluations with limited resolution and non-linearity show that ACM consistently improved upon the accuracy of BC while using the same hardware resources. The largest gain was obtained for non-linear weights when training with 5-bit precision XBar arrays or lower: effectively 2 bits in weight resolution were recovered, leading to a 20\% accuracy improvement. Compared to DE, ACM can achieve comparable training accuracies while reducing the read energy consumption by 7$\times$ and area by 2.3$\times$. Furthermore, the regularization effect of ACM makes it resilient to device variation. Assuming a 15\% device variation, ACM improves the inference accuracy of a VGG network trained on the CIFAR-10 dataset with 3-bit precision XBar arrays by an average of 10\% compared to other mappings.

\section*{Acknowledgment}
\footnotesize{This work was supported in part by ASCENT, one of six centers in JUMP, a SRC program sponsored by DARPA under task ID 2776.043.}

\begin{table}[tp]
\renewcommand{\arraystretch}{1.3}
\caption{System-level results of the three mapping approaches for training a two-layered MLP on XBar arrays.}
\label{tab:sys}
\centering
\begin{tabular}{|c|c|c|c|}
    \hline
    Mapping  &  BC & DE & ACM\\
    \hline
    \hline
    XBar Area ($\mu m^2$)& 914 & 2088 & 914 \\
    \hline
    Periphery Area ($\mu m^2$)& 157 & 246 & 157 \\
    \hline
    Read Energy ($\mu J$) & 2.402 & 14.408 & 2.402 \\
    \hline
    Read Delay ($ms$) & 0.240 & 0.318 & 0.240 \\
    \hline
\end{tabular}
\vspace{-0.2in}
\end{table}

{\footnotesize
\bibliography{bibfile}}
\bibliographystyle{ieeetr}

\end{document}